\documentclass[%
 reprint,
amsmath,amssymb,
aps,
]{revtex4-2}

\usepackage[pdftex]{graphicx}

\graphicspath{
{./Figures/},
}

\usepackage{diagbox}
\usepackage{braket}
\usepackage{dcolumn}
\usepackage{bm}
\usepackage{booktabs}
\usepackage{threeparttable}

\begin{document}

\preprint{APS/123-QED}

\title{Mathematical Crystal Chemistry}

\author{Ryotaro Koshoji}
\altaffiliation[Present adress:]{National Institute of Advanced Industrial Science and Technology (AIST), Nagoya, 463-8560, Japan} \email{r.koshoji@aist.go.jp}
\affiliation{Institute for Solid State Physics, The University of Tokyo, Kashiwa 277-8581, Japan}
\author{Taisuke Ozaki}%
\email{t-ozaki@issp.u-tokyo.ac.jp}
\affiliation{Institute for Solid State Physics, The University of Tokyo, Kashiwa 277-8581, Japan}

\date{\today}

\begin{abstract}

\noindent Efficient heuristics have predicted many functional materials such as high-temperature superconducting hydrides, while inorganic structural chemistry explains why and how the crystal structures are stabilized. Here we develop the paired mathematical programming formalism for searching and systematizing the structural prototypes of crystals. The first is the minimization of the volume of the unit cell under the constraints of only the minimum and maximum distances between pairs of atoms. We show the capabilities of linear relaxations of inequality constraints to optimize structures by the steepest-descent method, which is computationally very efficient. The second is the discrete optimization to assign five kinds of geometrical constraints including chemical bonds for pairs of atoms. Under the constraints, the two object functions, formulated as mathematical programming, are alternately optimized to realize the given coordination numbers of atoms. This approach successfully generates a wide variety of crystal structures of oxides such as spinel, pyrochlore-$\alpha$, and $\mathrm{K}_2 \mathrm{NiF}_4$ structures.

\end{abstract}

\maketitle

Discovering advanced materials enables technological innovation in many areas such as batteries, photovoltaics, and catalysis. Since electronic structure depends on the spatial arrangements of atoms, predictions of unknown crystal structures have a major impact on the discovery of better materials. However, computational structure searching is still difficult despite the progress of computational power, since structure prediction is a large-scale nonconvex minimization problem of the total energy, which has a large number of local minima, concerning the atomic arrangement and crystal lattice.

Finding the global minimum essentially involves visiting every local minimum, but some heuristic approaches with more or less intelligent ways such as evolutionally algorithms~\cite{10.1039/9781788010122, doi:10.1063/1.2210932, LYAKHOV20131172, Liu2021} and particle swarm-intelligence approach~\cite{PhysRevB.82.094116, WANG20122063, WANG2016406} have predicted novel materials including binary hydrides that have been experimentally synthesized as high-temperature superconductors under high pressure~\cite{PhysRevLett.119.107001, Liu6990, doi:10.1002/anie.201709970, Drozdov2019, PhysRevLett.122.027001, doi:10.1063/1.5128736, https://doi.org/10.1002/adma.202006832}. The drawbacks of these methods are that there are no guarantees that the most stable structure are found, and similarly, they can miss synthesizable unique structures. Furthermore, they cannot explain why and how the predicted structures are stabilized. This is also true of the deep learning approach~\cite{Pyzer-Knapp2022, Merchant2023, zeni2024mattergen, chen2024accelerating} and the random structure searching methods~\cite{PhysRevLett.97.045504, Pickard_2011}, while the latter may lead to peculiar structures without a \textit{priori} knowledge.

The integer programming approach has been developed to deterministically find the lowest energy structure with guarantees~\cite{Gusev2023}. The global optimum of integer programming can be found deterministically by brute force with branch-and-cut algorithms that are capable of rapidly eliminating large parts of the optimization domain from consideration. To apply this universal method, all the atoms are placed at discrete positions within a fixed cubic unit cell, and the total energy is represented by the sum of two-body classical interaction potentials~\cite{Gusev2023}. This approach has successfully found complex crystals such as garnet structure, but the fixed unit cells and approximated total energies may limit its applicability. Instead of using approximated total energies, geometrical approaches may be effective to search crystal structures, for example, Yokoyama \textit{et al.} have developed the dual periodic graph approach to generate structures composed of space-filling polyhedra~\cite{doi:10.1021/acs.cgd.3c01492}.

Important structural principles for ionic crystals have been summarized by Pauling in the five rules~\cite{doi:10.1021/ja01379a006, InorganicStructuralChemistry, StructuralInorganicChemistry}. George \textit{et al.}~have shown the limited predictive power of the Pauling rules since a few oxides simultaneously satisfy them~\cite{https://doi.org/10.1002/anie.202000829}, but the rules offer the basic concepts for inorganic structural chemistry. For example, the crystal structures of oxides, chalcogenides, and mixed anion compounds are linked polyhedra with explicit chemical bonds, especially between anions and cations~\cite{InorganicStructuralChemistry, StructuralInorganicChemistry}. Anions such as oxygen prefer to constitute the Barrow packings, which are the densest packings of spheres~\cite{10.2307/20159940}, and small cations are placed in the tetrahedral and octahedral sites, while large cations typically have a coordination number greater than 6, and in some cases, constitutes the Barrow packings with anions~\cite{InorganicStructuralChemistry, StructuralInorganicChemistry}. Small and large cations generally behave as hard and soft spheres, respectively. Highly polar bonds do not favor edge-sharing polyhedra and especially face-sharing polyhedra because of the increased electrostatic repulsion between central atoms~\cite{InorganicStructuralChemistry, StructuralInorganicChemistry}.

The structure of amorphous semiconductors is well represented by continuous random network models which require a condition that each atom should satisfy fully its bonding needs~\cite{WOS:000188444600058, WOS:A1985AED8200015, WOS:A1987M258200001, WOS:000089003200034}. Two atoms interact only if they are explicitly bonded, and the total energy is given by the Keating potential~\cite{WOS:A19667711700036, WOS:000089003200034}. The network is reproduced periodically by bond transpositions accepted with the Metropolis acceptance probability, conserving four coordination. This approach can generate structurally and electronically good structures comparable with experiments from random initial structures.

In this study, we propose a paired mathematical programming approach that finds candidate structures of crystals satisfying a series of empirical rules. The first is the minimization problem of the volume of the unit cell within the constraints of minimum and maximum distances between pairs of atoms. The constraints represent linked polyhedra and simultaneously the packings of anionic spheres regardless of the kinds of chemical bonds. We approximate the inequality constraints as the linear potentials for the structural relaxation. The second is the maximization problem of the number of chemical bonds. The five kinds of geometrical constraints, which impose restrictions on the distances between every pair of atoms, are assigned by the network of chemical bonds. The two optimization problems are alternately solved to find a proper structure where every cation satisfies its bonding needs; the transpositions of geometrical constraints transform random initial structures into linked polyhedra as the \textit{sillium} approach for amorphous silicons~\cite{WOS:000188444600058, WOS:A1985AED8200015, WOS:A1987M258200001, WOS:000089003200034}. We define a feasible solution as a structure satisfying all the constraints in the two optimization problems and all the bonding needs of atoms, and the other cases are denoted as infeasible solution. An optimal solution in this study is defined to be the feasible solution whose volume of the unit cell is minimized. The paired mathematical programming we propose is an efficient method which tries to find the optimal solution for a model system with a given composition.

Let us start our discussion by introducing the first object function in the paired mathematical programming formalism. The geometrical constraints are composed of the minimum and maximum distances between pairs of atoms. Suppose $\bm{x}_i$ is the cartesian coordinate of atom $i$, and $x_{ij \bm{T}}$ is the distances between atoms $i$ and $j\bm{T}$ given by
\begin{equation}
x_{ij \bm{T}} = \left|\bm{x}_j + \bm{T} - \bm{x}_i \right|,
\end{equation}
where the atom $j\bm{T}$ corresponds to the atom $j$ with the translational vector $\bm{T}$ of the crystal lattice. The object function is the volume $\Omega$ of the unit cell. The mathematical programming is given by
\begin{equation}
\begin{split}
\text{minimize} \quad & \Omega \\
\text{subject to} \quad & x_{ij \bm{T}} \ge d_{ij \bm{T}} ^{\left(\mathrm{min} \right)} \\
& x_{ij \bm{T}} \le d_{ij \bm{T}} ^{\left(\mathrm{max} \right)}
\end{split},
\label{eq:geometrical_optimization_problem}
\end{equation}
where $d_{ij \bm{T}} ^{\left(\mathrm{min} \right)}$ and $d_{ij \bm{T}} ^{\left(\mathrm{max} \right)}$ are the minimum and maximum distances between atoms $i$ and $j\bm{T}$, respectively. This model finds structural candidates of crystals without calculating the total energies.

To solve the problem, the equality and inequality constraints are relaxed as follows: The constraint of minimum distance between atoms $i$ and $j \bm{T}$:
\begin{equation}
x_{ij \bm{T}} \ge d_{ij \bm{T}} ^{\left(\mathrm{min} \right)},
\end{equation}
is relaxed to the hard-spherical potential $U_{\mathrm{min}} \left(x_{ij \bm{T}} \right)$ defined to be
\begin{equation}
U_{\mathrm{min}} \equiv
\begin{cases}
- k_{\downarrow} \left(x_{ij \bm{T}} - d_{ij \bm{T}} ^{\left(\mathrm{min} \right)} \right) & \text{$x_{ij \bm{T}} < d_{ij \bm{T}} ^{\left(\mathrm{min} \right)}$} \\
0 & \text{$d_{ij \bm{T}} ^{\left(\mathrm{min} \right)} \le x_{ij \bm{T}}$}
\end{cases}, \label{eq:minimum_constraint_potential}
\end{equation}
where $k_{\downarrow}$ is a common constant for the minimum constraints. Similarly, the constraint of maximum distance between atoms $i$ and $j \bm{T}$:
\begin{equation}
x_{ij \bm{T}} \le d_{ij \bm{T}} ^{\left(\mathrm{max} \right)},
\end{equation}
is relaxed to the hard-spherical potential $U_{\mathrm{max}} \left(x \right)$ defined to be
\begin{equation}
U_{\mathrm{max}} \equiv
\begin{cases}
0 & \text{$x_{ij \bm{T}} \le d_{ij \bm{T}} ^{\left(\mathrm{max} \right)}$} \\
k_{\uparrow} \left(x_{ij \bm{T}} - d_{ij \bm{T}} ^{\left(\mathrm{max} \right)} \right) & \text{$d_{ij \bm{T}} ^{\left(\mathrm{max} \right)} < x_{ij \bm{T}}$}
\end{cases}, \label{eq:maximum_constraint_potential}
\end{equation}
where $k_{\uparrow}$ is a common constant for the maximum constraints. The relaxations transform the problem of Eq.~\eqref{eq:geometrical_optimization_problem} into the minimization problem of the enthalpy $H \left(\left\{\bm{x}_i , \bm{a}_i \right\} \right)$ per unit cell formulated as
\begin{equation}
\text{minimize} \quad H \left(\left\{\bm{x}_i , \bm{a}_i \right\} \right) = E \left(\left\{\bm{x}_i , \bm{a}_i \right\} \right) + P \Omega,
\label{eq:approximated_geometrical_optimization_problem}
\end{equation}
where $P$ is the pressure and $E \left(\left\{\bm{x}_i , \bm{a}_i \right\} \right)$ is given by
\begin{equation}
E \left(\left\{\bm{x}_i, \bm{a}_i \right\} \right) = \sum_{i \le j} \sum_{\bm{T}} \left[U_{\mathrm{min}} \left(x_{ij\bm{T}} \right) + U_{\mathrm{max}} \left(x_{ij\bm{T}} \right) \right],
\end{equation}
and $\bm{a}_i$ is the primitive translation vectors of crystal lattice. The displacement of each atom is calculated from the derivative of the enthalpy $H$ per unit cell as
\begin{equation}
\Delta \bm{x}_i = - k_i \frac{\mathrm{d} H}{\mathrm{d} \bm{x}_i}.
\end{equation}
The constant $k_i$ is scaled to ensure so that the norm of displacement does not exceed the given maximum value, $\Delta x_{\mathrm{max}}$, for each optimization step. Similarly, the primitive lattice vectors $\bm{a}_i$ is optimized as
\begin{equation}
\Delta \bm{a}_i = - k_{\mathrm{L}} \frac{\mathrm{d} H}{\mathrm{d} \bm{a}_i},
\end{equation}
where $k_{\mathrm{L}}$ is set to satisfy the condition:
\begin{equation}
\sqrt{\left| \Delta \bm{a}_1 \right|^2 + \left| \Delta \bm{a}_2 \right|^2 + \left| \Delta \bm{a}_3 \right|^2} \le \Delta A_{\mathrm{max}}.
\end{equation}
To reach the local optima of the problem given by Eq.~\eqref{eq:approximated_geometrical_optimization_problem}, we gradually reduce $\Delta x_{\mathrm{max}}$ and $\Delta A_{\mathrm{max}}$ as
\begin{equation}
\begin{split}
\Delta x_{\mathrm{max}} ^{\left( n \right)} &= \alpha ^{n-1} \Delta x_{\mathrm{max}} ^{\left( 1 \right)}, \\
\Delta A_{\mathrm{max}} ^{\left( n \right)} &= \frac{\Delta x_{\mathrm{max}} ^{\left( n \right)}}{4000} \sum_i \frac{4 \pi}{3} \left(r_i ^{\left(\mathrm{I} \right)} \right)^3 ,
\end{split}
\end{equation}
where $\alpha$ is a constant and $n$ is the number of geometrical optimization steps. The values of pameters such as $\alpha$,  $k_{\uparrow}$, $k_{\downarrow}$, and $P$ are discussed in Supplementary Information~\cite{SupplementalMaterial}. One may consider that structures are never relaxed due to the discontinuity in the first derivative of the potential. However, the structure converges to an optimal solution by gradually reducing the maximum $\Delta x_{\mathrm{max}}$ and $\Delta A_{\mathrm{max}}$, even though there is the discontinuity. The effectiveness of linear relaxations of inequality constraints has already been shown in the previous studies on the densest sphere packings as iterative-balance method~\cite{PhysRevE.103.023307, PhysRevE.104.024101, koshoji2021diverse}: The method enables the structures to reach a local optima precisely enough to calculate packing fractions. The inequality constraints are widely approximated by the logarithmic barrier functions~\cite{ConvexOptimization}, but the advantage of the linear potentials consists of the two folds: One is that the computational cost is the lowest and the other is that the potentials can impose hard penalties only when the constraints are not satisfied.

\begin{figure}
\centering
\includegraphics[width=0.6\columnwidth]{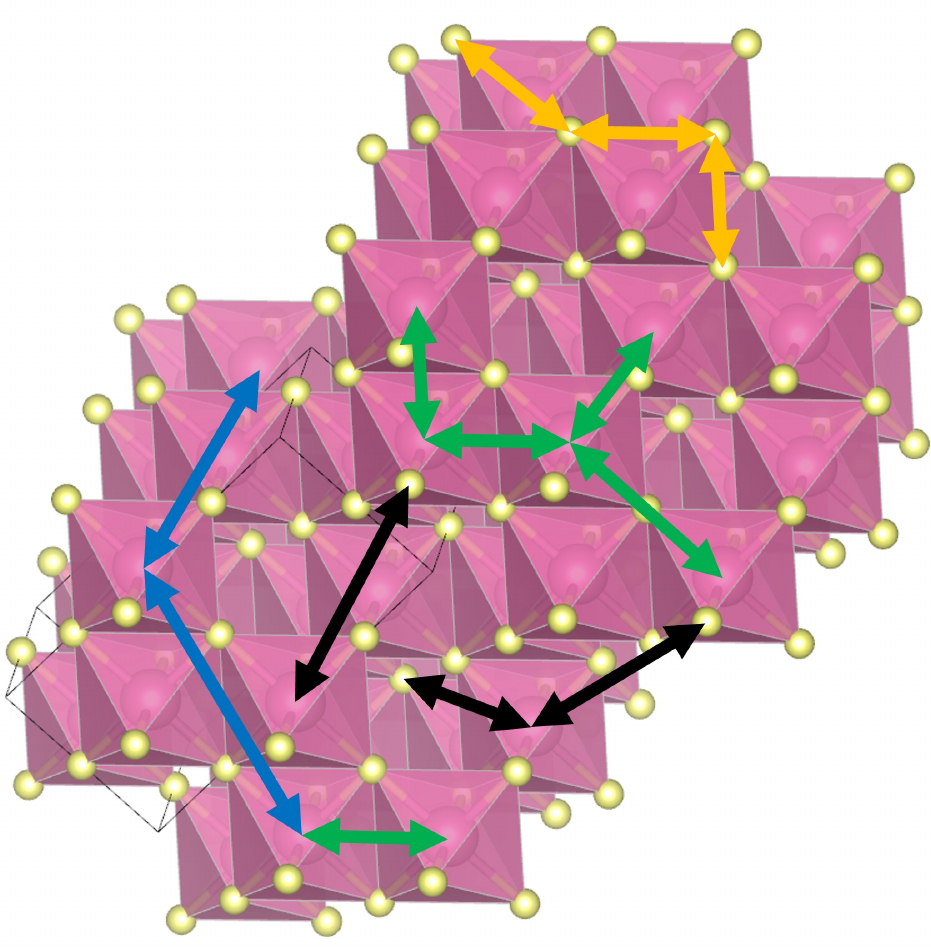}
\caption{
The five kinds of geometrical constraints with the optimal solutions of $\mathrm{Hd}_4 \mathrm{O}_8$ corresponding to anatase structure.
Cylinders connecting to two atoms correspond to chemical bonds (CBs).
Black, orange, blue, and green arrows correspond to non-bonding constraints (NBCs), anionic constraints (ACs), cationic constraints (CCs), and polyhedral constraints (PCs), respectively.
}
\label{fig:geometrical_constraints}
\end{figure}

\begin{table*}
\caption{The list of geometrical constraints. If a constraint is inactive, the mimimum and maximum distances are $0$ and $\infty$, respectively. $r ^{\left(\mathrm{I} \right)}$ and $R ^{\left(\mathrm{I} \right)}$ are minimum and maximum ionic radii, respectively. $r ^{\left(\mathrm{C} \right)}$ are cationic repulsion radii.}
\label{table:geometrical_constraints}
\begin{ruledtabular}
\begin{tabular}{cccc}
Constraint type & Abbreviation & Minimum distance if active & Maximum distance if active \\ \hline
Chemical Bond & CB & $d_{ij \bm{T}} ^{\left(\mathrm{CB}, \mathrm{min} \right)} = r_i ^{\left(\mathrm{I} \right)} + r_j ^{\left(\mathrm{I} \right)}$ & $d_{ij \bm{T}} ^{\left(\mathrm{CB}, \mathrm{max} \right)} = R_i ^{\left(\mathrm{I} \right)} + R_j ^{\left(\mathrm{I} \right)}$ \\
Non-Bonding Constraint & NBC & $d_{ij \bm{T}} ^{\left(\mathrm{NBC}, \mathrm{min} \right)} = 1.2 \left(R_i ^{\left(\mathrm{I} \right)} + R_j ^{\left(\mathrm{I} \right)} \right)$ & $d_{ij \bm{T}} ^{\left(\mathrm{NBC}, \mathrm{max} \right)} = \infty$ \\
Anionic Constraint & AC & $d_{ij \bm{T}} ^{\left(\mathrm{AC}, \mathrm{min} \right)} = R_i ^{\left(\mathrm{I} \right)} + R_j ^{\left(\mathrm{I} \right)}$ & $d_{ij \bm{T}} ^{\left(\mathrm{AC}, \mathrm{max} \right)} = \infty$ \\
Cationic Constraint & CC & $d_{ij \bm{T}} ^{\left(\mathrm{CC}, \mathrm{min} \right)} = r_i ^{\left(\mathrm{C} \right)} + r_j ^{\left(\mathrm{C} \right)}$ & $d_{ij \bm{T}} ^{\left(\mathrm{CC}, \mathrm{max} \right)} = \infty$ \\
Polyhedral Constraint & PC & Depending on $N_{ij}^{\left(\mathrm{CBA} \right)}$ as discussed in main text & $d_{ij \bm{T}} ^{\left(\mathrm{PC}, \mathrm{max} \right)} = \infty$
\end{tabular}
\end{ruledtabular}
\end{table*}

Next we turn our discussion to the second object function in the paired mathematical programming formalism. The problem is how to determine the values of $d_{ij \bm{T}} ^{\left(\mathrm{min} \right)}$ and $d_{ij \bm{T}} ^{\left(\mathrm{max} \right)}$. In the following discussion, we focus on the crystal structures of oxides. We introduce the five kinds of geometrical constraints as listed in Table \ref{table:geometrical_constraints} and illustrated in Fig.~\ref{fig:geometrical_constraints}. The minimum and maximum distances are determined as
\begin{equation}
\begin{split}
d_{ij \bm{T}} ^{\left(\mathrm{min} \right)} &= \max_{\mathrm{p}} \left\{ s_{ij \bm{T}} ^{\left(\mathrm{p} \right)} \, d_{ij \bm{T}} ^{\left(\mathrm{p, min} \right)} \right\} \\
d_{ij \bm{T}} ^{\left(\mathrm{max} \right)} &= \max_{\mathrm{p}} \left\{ s_{ij \bm{T}} ^{\left(\mathrm{p} \right)} \, d_{ij \bm{T}} ^{\left(\mathrm{p, max} \right)} \right\} \\
\sum_{p} s_{ij \bm{T}} ^{\left(\mathrm{p} \right)} &= 1
\end{split},
\end{equation}
where $s_{ij \bm{T}} ^{\left(\mathrm{p} \right)} \in \left\{ 0,1 \right\}$ is a switch variable selecting one of the geometrical constraints, and $p$ runs chemical bond (CB), non-bonding constraint (NBC), anionic constraint (AC), cationic constraint (CC), and polyhedral constraint (PC). The geometrical constraints for every pair of atoms are assigned depending on the network of CBs. CBs or NBCs are formed between every pair of an anion and a cation. ACs are formed between every pair of anions. Besides, if two coordination polyhedra around cations share some bridging anions, PC is formed between the two cations. Finally, CCs are formed between remained pairs of cations. To assist making coordination polyhedra, NBCs spatially separate the anion and cation when CB is not formed between them. ACs impose the hard-spherical constraint for every pair of anions; oxygen ions often constitute the Barrow-packings if the cationic radii are small. CCs are necessary to spatially separate two cations enough to avoid the condensation of cations, while a minimum distance of PC is smaller than that of CC to share a common edge or face as
\begin{equation}
d_{ij \bm{T}} ^{\left(\mathrm{PC}, \mathrm{min} \right)} =
\begin{cases}
r_i ^{\left(\mathrm{M} \right)} + r_j ^{\left(\mathrm{M} \right)} & \text{$N_{ij} ^{\left(\mathrm{CBA} \right)} = 1$} \\
D_{i}^{\left(\text{shared} \right)} + D_{j}^{\left(\text{shared} \right)} & \text{otherwise}
\end{cases},
\end{equation}
where $D_{i} ^{\left(\text{shared} \right)}$ is the minimum distance from the cation $i$ to the shared common edge or face as detailed in Supplementary Information~\cite{SupplementalMaterial}. 

Since any cation $i$ has the desirable coordination number $N_i ^{\left(\mathrm{CB} \right)}$ for CBs, the feasible linked-polyhedra must satisfy the conditions of
\begin{equation}
s_i ^{\left( + \right)} \left( N_{i} ^{\left(\mathrm{CB} \right)} - n_{i} ^{\left(\mathrm{CB} \right)} \right) = 0,
\label{eq:coordination_number_condition}
\end{equation}
where $n_{i} ^{\left(\mathrm{CB} \right)}$ is the coordination number, which is counted by the number of CBs for atom $i$, and $s_i ^{\left( + \right)}$ is a switch function for the atom $i$ given by
\begin{align}
s_i ^{\left( + \right)} =
\begin{cases}
1 & \text{if the atom $i$ is cation} \\
0 & \text{if the atom $i$ is anion}
\end{cases}.
\end{align}
Most of random initial structures are infeasible solutions, and the analytic optimization methods such as the steepest-descent method cannot escape from the infeasible solution to a feasible solution. Therefore, we periodically update the values of $d_{ij \bm{T}} ^{\left(\mathrm{min} \right)}$ and $d_{ij \bm{T}} ^{\left(\mathrm{max} \right)}$ depending on the network of CBs that are also renewed periodically. The network is determined from the maximization problem of the number of CBs given by
\begin{equation}
\begin{split}
\text{maximize} \quad & s_1 ^{\left( + \right)} \varepsilon_1 n_{1} ^{\left(\mathrm{CB} \right)} + \cdots + s_M ^{\left( + \right)} \varepsilon_M n_{M} ^{\left(\mathrm{CB} \right)} \\
\text{subject to} \quad & n_{i} ^{\left(\mathrm{CB} \right)} \le N_{i} ^{\left(\mathrm{CB} \right)} \\
& n_{ij \bm{T}} ^{\left(\mathrm{CBA} \right)} \le N_{ij} ^{\left(\mathrm{CBA} \right)} \\
& s_{ij \bm{T}} ^{\left(\mathrm{CB} \right)} x_{ij \bm{T}} \le 2 d_{ij \bm{T}} ^{\left(\mathrm{CB}, \mathrm{max} \right)} \\
& \sum _{j, \bm{T}} s_i ^{\left( + \right)} s_{ij \bm{T}} ^{\left(\mathrm{CB} \right)} x_{ij \bm{T}} \le \min \sum_{q=1} ^{n_i} x_{i j_q \bm{T}_q}
\end{split}
\label{eq:chemical-bonding_optimization_problem}
\end{equation}
where $0 \le \varepsilon _i$ is the fixed bonding affinity, $n_{ij \bm{T}} ^{\left(\mathrm{CBA} \right)}$ is the number of common bridging atoms between the pair of atoms $i$ and $j \bm{T}$, $N_{ij} ^{\left(\mathrm{CBA} \right)}$ is the default maximum number of common bridging atoms defined in Eq.~\eqref{eq:max_common_bridging_atoms}, and $q$ is an index for selecting $n_i$ indices of anions $j \bm{T}$ randomly. The third constraints define the formation ranges of CBs. The fourth constraints force cations to create CBs with the nearest $n_i$ anions. The optimization problem can be solved as follows: First, all the cations create as many CBs as possible with neighboring anions, and seconds, we erase CBs with maximizing the object function of Eq.~\eqref{eq:chemical-bonding_optimization_problem} until all the number of common bridging atoms satisfy the second constraints. Note that two polyhedra are linked by sharing a common vertex, a common edge, or a common face that corresponds to sharing one, two, or more than two common bridging atoms, respectively~\cite{InorganicStructuralChemistry}.

\begin{figure}
\centering
\includegraphics[width=1.0\columnwidth]{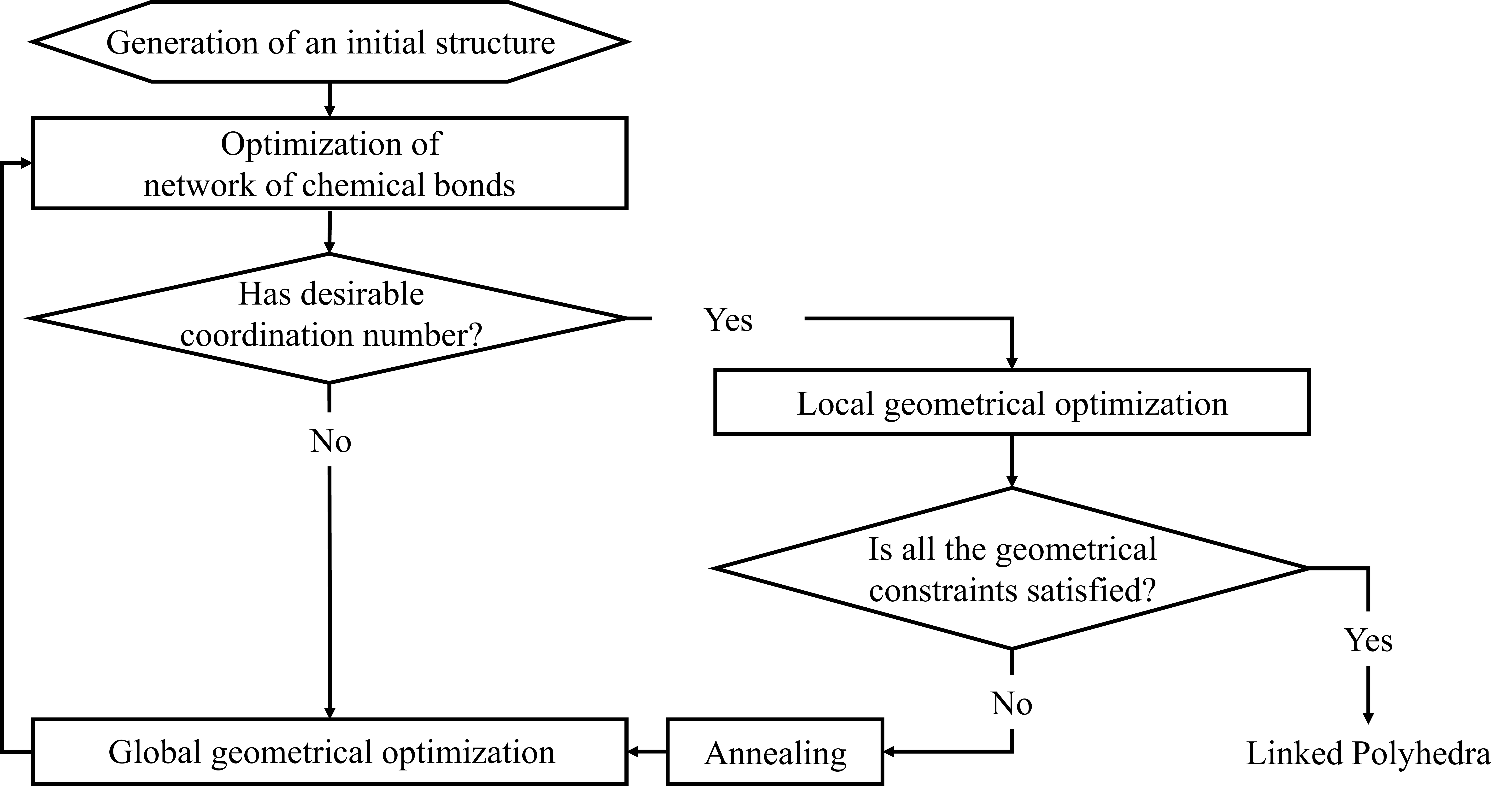}
\caption{
The flow chart of the optimization scheme to find the optimal solutions of the paired mathematical programming.
}
\label{fig:flowChart}
\end{figure}

To find the optimal solutions, we solve the problems of Eqs.~\eqref{eq:geometrical_optimization_problem} and \eqref{eq:chemical-bonding_optimization_problem} alternately. The optimization scheme is shown in Fig.~\ref{fig:flowChart}. A random initial structure can be infeasible solution, in fact, not all cations may not be surrounded by oxygen ions and may not have desirable coordination numbers. Besides, even if all the atoms have the desirable coordination numbers, the structure does not necessarily satisfy all the geometrical constraints such as the maximum number of common bridging atoms. Therefore, the global geometrical optimization is aimed at transforming the structure largely enough to create a different network of CBs by large $\Delta x_{\mathrm{max}}$. After the small number of global geometrical optimization steps, we solve the problem of Eq.~\eqref{eq:chemical-bonding_optimization_problem} once again to update the network of CBs. The mutable network and the other geometrical constraints assist cations making coordination polyhedra, and accordingly, the structure is transformed into a feasible solution. If the structure satisfies the condition of Eq.~\eqref{eq:coordination_number_condition}, the structure is locally optimized by small $\Delta x_{\mathrm{max}}$ to identify whether the structure can be the optimal solusion of the geometrical optimization problem given by Eq.~\eqref{eq:geometrical_optimization_problem}. If so, the structure is the optimal solution of the paired optimization problems. If not, the structure is globally optimized again after annealing. The parameters of global and local optimization are detailed in Supplementary Information~\cite{SupplementalMaterial}.

To estimate the number of optimal solutions and coarsely classify the optimal solutions, we define a structural fingerprint as the bundle of the fingerprints by the lists of geometrical constraints (LGCs) in dictionary order. LGCs are defined as what kinds of geometrical constraints are formed with the adjacent atoms. The fingerprint of LGC enumerates the element symbols of the adjacent atoms linked by CBs, PCs, and ACs in dictionary order, respectively. Note that there is a possibility that the same structure fingerprint is assigned to different structures. For example, the fingerprint for the rutile structure and $\alpha$-$\mathrm{PbO}_2$ is the case. See also Supplementary Information~\cite{SupplementalMaterial}.

\begin{table}
\caption{11 kinds of cations. $C_n$ ($n \in \mathbb{N}$) corresponds to the minimum radius so that the cation can connect $n$ oxygen ions, and $C_{O}$ corresponds to the ionic radius of oxygen ion. The values of the radii are listed in Supplementary Information~\cite{SupplementalMaterial}. A symbol represents the atomic radius and/or the maximum number of common bridging atoms. For example, the minimum and maximum cationic radii of the cation $\mathrm{Op}$ are $C_7$ (sePta) and $C_8$ (Octa), respectively, while the maximum number of common bridging atoms $\mathrm{Ht}$ is three (Tri). The color is given if the cation is illustrated in Figures.}
\label{table:cations}
\begin{ruledtabular}
\begin{tabular}{cccccc}
Symbol & $r^{\left(\mathrm{I} \right)}$ & $R^{\left(\mathrm{I} \right)}$ & $N^{\left(\mathrm{CB} \right)}$ & $N ^{\left(\mathrm{CBA} \right)}$ & color \\ \hline
T & $C_{4}$ & $C_{4}$ & 4 & 1 & sky blue \\
Pe & $C_{5}$ & $C_{5}$ & 5 & 2 & - \\
Hd & $C_{6}$ & $C_{6}$ & 6 & 2 & rose \\
Ht & $C_{6}$ & $C_{6}$ & 6 & 3 & red \\
Sh & $C_{7}$ & $C_{7}$ & 6 & 2 & - \\
Op & $C_{7}$ & $C_{8}$ & 7 & 4 & dark blue \\
Eo & $C_{8}$ & $C_{9}$ & 8 & 4 & purple \\
Uo & $C_{7}$ & $C_{11}$ & 8 & 4 & blue \\
Do & $C_{7}$ & $C_{O}$ & 8 & 4 & - \\
De & $C_{O}$ & $C_{O}$ & 9 & 4 & light green \\
D & $C_{O}$ & $C_{O}$ & 12 & 4 & green
\end{tabular}
\end{ruledtabular}
\end{table}

The paired mathematical programming approach is aimed at showcasting capabilities of simple rules to search structural prototypes of crystals. In this study, we introduce $11$ kinds of cations shown in Table \ref{table:cations} to reproduce a wide variety of coordination polyhedra. The $11$ cations are selected by considering closed packed polyhedra consisting of the centered cation and surrounding oxygen ions, and defined by $r^{\left(\mathrm{I} \right)}$, $R^{\left(\mathrm{I} \right)}$, $N^{\left(\mathrm{CB} \right)}$, and $N ^{\left(\mathrm{CBA} \right)}$, where $N ^{\left(\mathrm{CBA} \right)}$ is the default maximum number of common bridging atoms. We determine $N_{ij}^{\left(\mathrm{CBA} \right)}$ in the second constraint of Eq.~\eqref{eq:chemical-bonding_optimization_problem} as
\begin{equation}
N_{ij}^{\left(\mathrm{CBA} \right)} = \min \left\{ N_{i}^{\left(\mathrm{CBA} \right)}, N_{j}^{\left(\mathrm{CBA} \right)} \right\}.
\label{eq:max_common_bridging_atoms}
\end{equation}
Cationic radius $C_n$ ($n \in \mathbb{N}$) corresponds to the minimum radius so that the cation can connect $n$ oxygen ions, while $C_{O}$ correponds to the ionic radius of oxygen ion. They are given in Supplementary Information~\cite{SupplementalMaterial}. If the minimum and maximum cationic radii are the same, the cation is a hard sphere; if not, the cation is a soft sphere, because the bonding lenths can vary within the two radii.

\begin{table*}
\caption{The number of the discovered optimal solutions for each composition and the confirmed correspondence of optimal solutions with crystal structures. Note that if the packing fraction is less than the minimum, $60 \%$, which is given in Table II of Supplementary Information~\cite{SupplementalMaterial}, we reject the structure from the optimal solutions in this study.}
\label{table:results}
\begin{ruledtabular}
\begin{tabular}{ccc}
Composition per unit cell & Number of discovered optimal solutions & Confirmed real crystal structures \\ \hline
$\mathrm{Ht}_2 \mathrm{T}_2 \mathrm{O}_{6}$ & 21 & $\mathrm{Ga}_2 \mathrm{O}_3$ \\
$\mathrm{Ht}_3 \mathrm{T}_5 \mathrm{O}_{12}$ & 508 & - \\
$\mathrm{Ht}_4 \mathrm{T}_4 \mathrm{O}_{12}$ & 597 & $\mathrm{Ga}_2 \mathrm{O}_3$ \\
$\mathrm{Ht}_4 \mathrm{T}_2 \mathrm{O}_{8}$ & 29 & Spinel \\
$\mathrm{Ht}_8 \mathrm{T}_4 \mathrm{O}_{16}$ & 216 & Spinel \\
$\mathrm{D}_3 \mathrm{Ht}_6 \mathrm{O}_{15}$ & 14 & $\mathrm{BaTi}_2 \mathrm{O}_5$ \\
$\mathrm{D}_4 \mathrm{Ht}_4 \mathrm{O}_{12}$ & 4 & Perovskite, $\mathrm{BaNiO}_3$, $\mathrm{BaMnO}_3$, $\mathrm{Cs}_2 \mathrm{NaCrF}_6$ \\
$\mathrm{D}_1 \mathrm{De}_2 \mathrm{Ht}_2 \mathrm{O}_{7}$ & 1 & $\mathrm{La}_{2-x} \mathrm{Sr}_{1+x} \mathrm{Cu}_2 \mathrm{O}_7$ \\
$\mathrm{D}_2 \mathrm{De}_4 \mathrm{Ht}_4 \mathrm{O}_{14}$ & 1 & $\mathrm{La}_{2-x} \mathrm{Sr}_{1+x} \mathrm{Cu}_2 \mathrm{O}_7$ \\
$\mathrm{De}_4 \mathrm{Ht}_2 \mathrm{O}_{8}$ & 1 & $\mathrm{La}_{2-x} \mathrm{Sr}_x \mathrm{CuO}_4$ \\
$\mathrm{De}_8 \mathrm{Ht}_4 \mathrm{O}_{16}$ & 1 & $\mathrm{La}_{2-x} \mathrm{Sr}_x \mathrm{CuO}_4$ \\
$\mathrm{Do}_4 \mathrm{Hd}_4 \mathrm{O}_{14}$ & 6 & Pyrochlore-$\alpha$ \\
$\mathrm{Uo}_4 \mathrm{Hd}_4 \mathrm{O}_{14}$ & 20 & Pyrochlore-$\alpha$ \\
$\mathrm{Eo}_4 \mathrm{Ht}_4 \mathrm{O}_{12}$ & 21 & - \\
$\mathrm{Eo}_6 \mathrm{O}_{12}$ & 944 & Zirconia \\
$\mathrm{Op}_4 \mathrm{Ht}_4 \mathrm{O}_{12}$ & 95 & $\mathrm{YFeO}_3$ \\
$\mathrm{Op}_4 \mathrm{Ht}_8 \mathrm{O}_{16}$ & 15 & - \\
$\mathrm{Pe}_2 \mathrm{Sh}_1 \mathrm{O}_{4}$ & 4 & $\mathrm{InGaZnO}_4$ \\
\end{tabular}
\end{ruledtabular}
\end{table*}

We apply our mathematical programming model to several compositions of cations, and find that a wide variety of real crystals are discovered in the optimal solutions by the paired mathematical programming. Table \ref{table:results} shows the number of discovered optimal solutions for each composition, which seems to be much smaller than the number of local minima in the total energy by \textit{ab-initio} simulations, and corresponding real crystals found in them. We discuss here results by focusing on aspects of the generated coordination polyhedra.

\begin{figure}
\centering
\includegraphics[width=1.0\columnwidth]{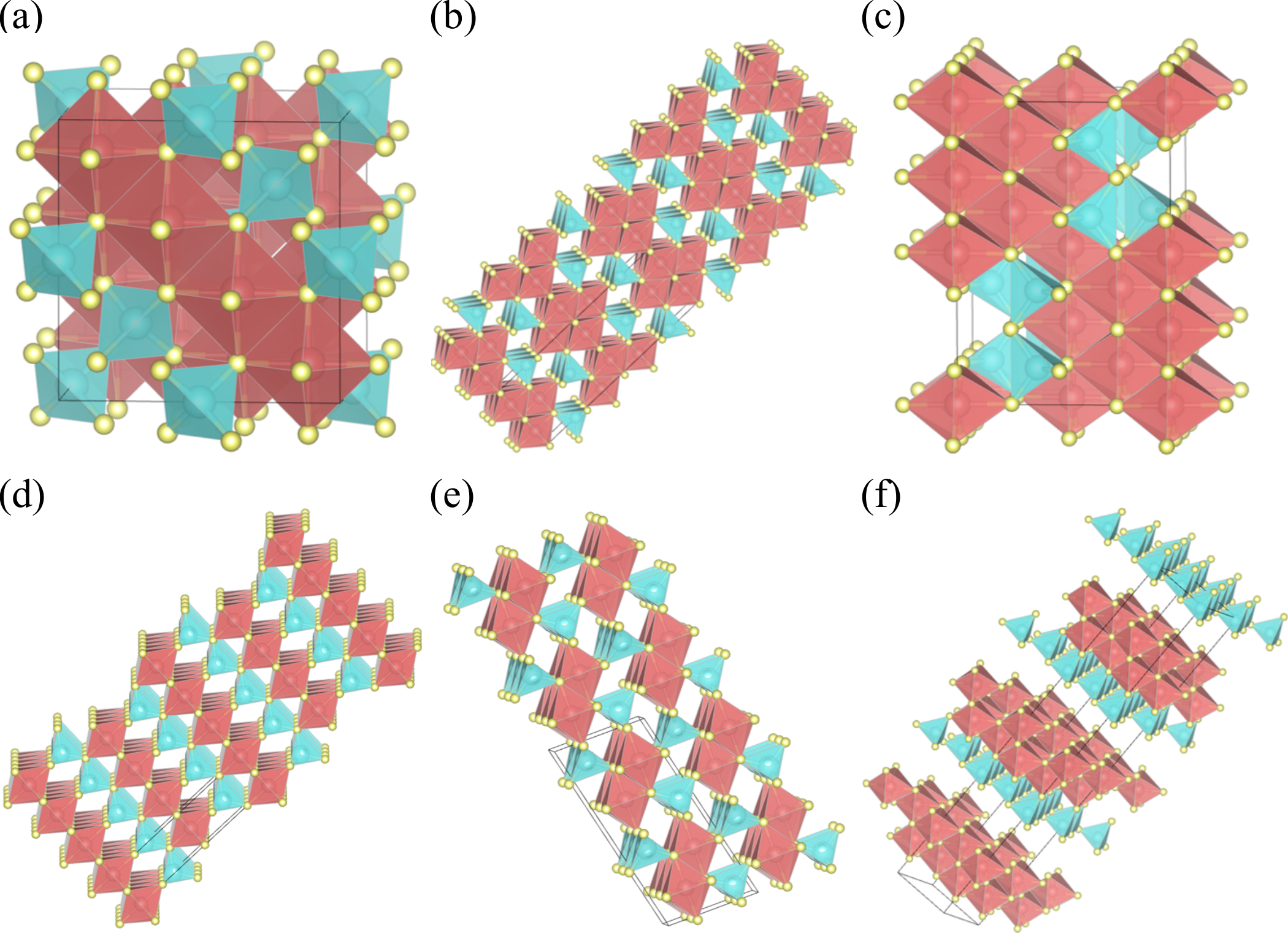}
\caption{
(a) $\mathrm{Ht}_8 \mathrm{T}_4 \mathrm{O}_{16}$ corresponding to the spinel structure (3 LGCs). Yellow balls correspond to oxygen ions.
(b) $\mathrm{Ht}_8 \mathrm{T}_4 \mathrm{O}_{16}$ with the $Cmmm$ symmetry (7 LGCs).
(c) $\mathrm{Ht}_8 \mathrm{T}_4 \mathrm{O}_{16}$ with the $Imma$ symmetry (6 LGCs). The structure is similar to the crystal of $\mathrm{Li}_2 \mathrm{WO}_{4}$ (6 LGCs). The difference comes from the sizes of the unit cells. 
(d) $\mathrm{Ht}_4 \mathrm{T}_4 \mathrm{O}_{12}$ with the $I \bar{4} m2$ symmetry (4 LGCs).
(e) $\mathrm{Ht}_4 \mathrm{T}_4 \mathrm{O}_{12}$ corresponding to the $\mathrm{Ga}_2 \mathrm{O}_{3}$ structure (5 LGCs).
(f) $\mathrm{Ht}_5 \mathrm{T}_3 \mathrm{O}_{12}$ with the $R3$ symmetry (7 LGCs).
The illustrated structures are symmetrized and their space groups are determined by the code \textsl{SPGLIB}~\cite{togo2018textttspglib} with correcting the structural distortions. All the structural figures in this study are generated by \textsl{VESTA}~\cite{Momma:db5098}.  
}
\label{fig:Hd-T-O}
\end{figure}

The optimal solutions of $\mathrm{Ht}_l \mathrm{T}_m \mathrm{O}_n$ form a wide variety of linked-polyhedra composed of tetrahedra and octahedra as shown in Fig.~\ref{fig:Hd-T-O}. The cations $\mathrm{T}$ and $\mathrm{Ht}$ are placed in the tetrahedral and octahedral sites, respectively, in the Barrow-packings consisting of oxygen ions. Many structures are rejected from the optimal solutions by the constraints of the maximum number of common bridging atoms of $\mathrm{T}$. The spinel structure has the lowest number of LGCs and highest symmetry, found as one of the optimal solutions for both the $\mathrm{Ht}_4 \mathrm{T}_2 \mathrm{O}_{8}$ and $\mathrm{Ht}_8 \mathrm{T}_4 \mathrm{O}_{16}$, while the $\mathrm{Ga}_2 \mathrm{O}_{3}$ structure, which is composed of 5 LGCs with the $C2/m$ symmetry, has more LGCs and lower symmetry than the structure shown in Fig.~\ref{fig:Hd-T-O}(d). Real crystal structures are not necessarily the optimal solution composed of the lowest number of LGCs and highest symmetry, however, they are generally good indicators to assess expectations of the structures to be realized by real crystals. The optimal solution shown in Fig.~\ref{fig:Hd-T-O}(c) is similar to the crystal structure of $\mathrm{Li}_2 \mathrm{WO}_4$; the difference comes from the sizes of the unit cells. The crystal structure of $\mathrm{Li}_2 \mathrm{WO}_4$ is composed of large tetrahedra and octahedra around lithiums, but this structure may be reproduced by using cations $\mathrm{Ht}$ and $T$, which are placed in the octahderal and tetrahedral sites in the Barrow-packings consisting of oxygens, as the optimal solution of $\mathrm{Ht}_{16} \mathrm{T}_{8} \mathrm{O}_{32}$. Figure \ref{fig:Hd-T-O}(f) shows the layered structure terminated by tetrahedra and octahedra. This optimal solution is composed of the lowest number of LGCs in $\mathrm{Ht}_5 \mathrm{T}_3 \mathrm{O}_{12}$.

\begin{figure}
\centering
\includegraphics[width=1.0\columnwidth]{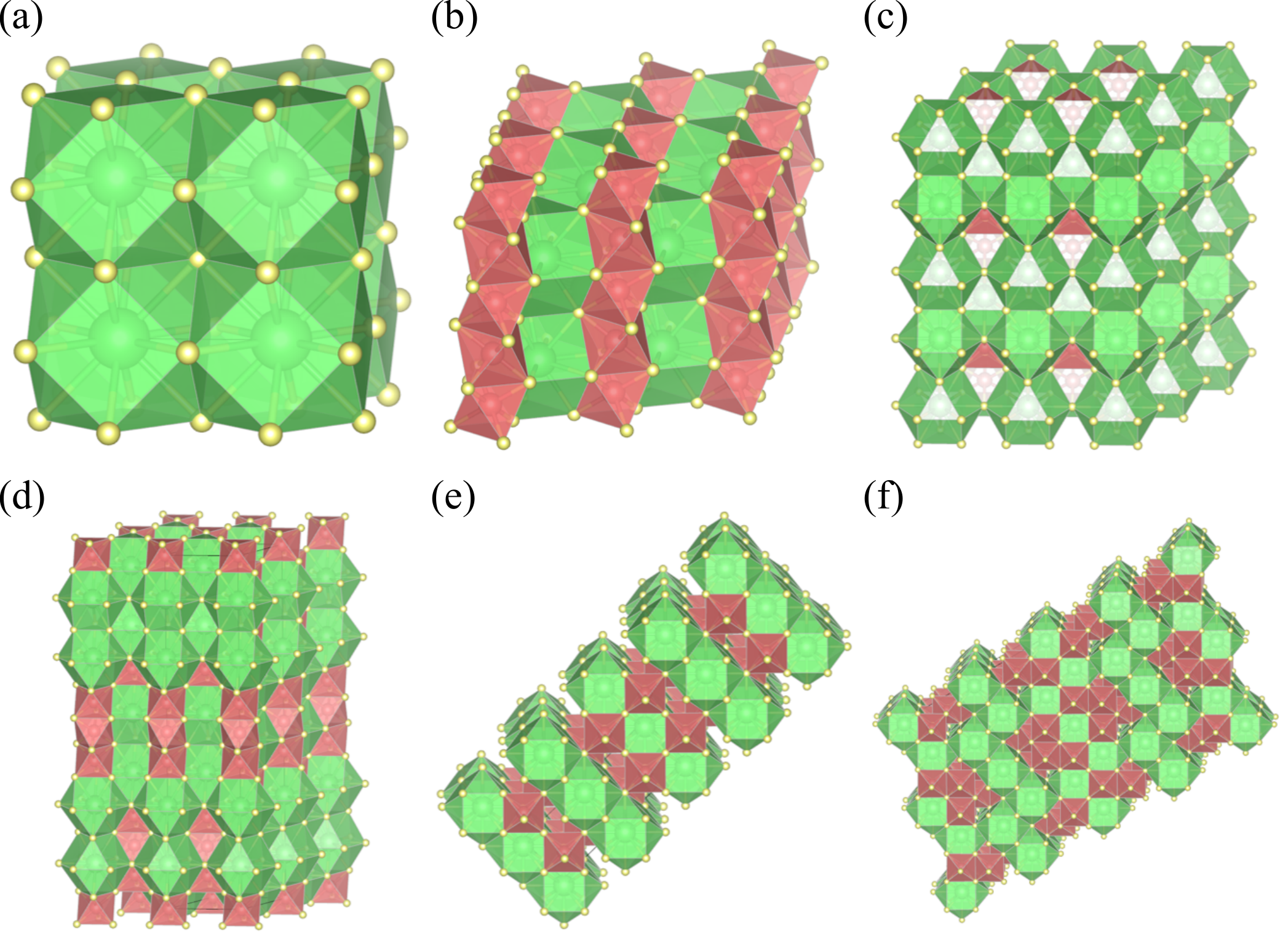}
\caption{
(a) $\mathrm{D}_4 \mathrm{Ht}_4 \mathrm{O}_{12}$ corresponding to the perovskite structure (3 LGCs).
(b) $\mathrm{D}_4 \mathrm{Ht}_4 \mathrm{O}_{12}$ corresponding to the crystal of $\mathrm{BaNiO}_{3}$ (3 LGCs).
(c) $\mathrm{D}_4 \mathrm{Ht}_4 \mathrm{O}_{12}$ corresponding to the crystal of $\mathrm{BaMnO}_{3}$ (4 LGCs).
(d) $\mathrm{D}_4 \mathrm{Ht}_4 \mathrm{O}_{12}$ corresponding to the crystal of $\mathrm{Cs}_2 \mathrm{NaCrO}_{6}$ (5 LGCs).
(e) $\mathrm{D}_3 \mathrm{Ht}_6 \mathrm{O}_{15}$ with the $Cmmm$ symmetry (5 LGCs).
(f) $\mathrm{D}_3 \mathrm{Ht}_6 \mathrm{O}_{15}$ corresponding to the crystal of $\mathrm{BaTi}_2 \mathrm{O}_{5}$ (11 LGCs).
}
\label{fig:9_and_12_coordination}
\end{figure}

All the optimal solutions of $\mathrm{D}_4 \mathrm{Ht}_4 \mathrm{O}_{12}$ correspond to the real crystals shown in Figs.~\ref{fig:9_and_12_coordination}(a), \ref{fig:9_and_12_coordination}(b), \ref{fig:9_and_12_coordination}(c), and \ref{fig:9_and_12_coordination}(d). The cation $\mathrm{D}$ makes coordination polyhedra of anticuboctahedron or cuboctahedron. Besides, we find that $\mathrm{D}_3 \mathrm{Ht}_6 \mathrm{O}_{15}$ has the optimal solution corresponding to the $\mathrm{BaTi}_2 \mathrm{O}_5$ structure shown in Fig.~\ref{fig:9_and_12_coordination}(f). The optimal solution consists of 11 LGCs, but we find another optimal solution consiting of only $5$ LGCs with the $Cmmm$ symmetry shown in Fig.~\ref{fig:9_and_12_coordination}(e).

\begin{figure}
\centering
\includegraphics[width=1.0\columnwidth]{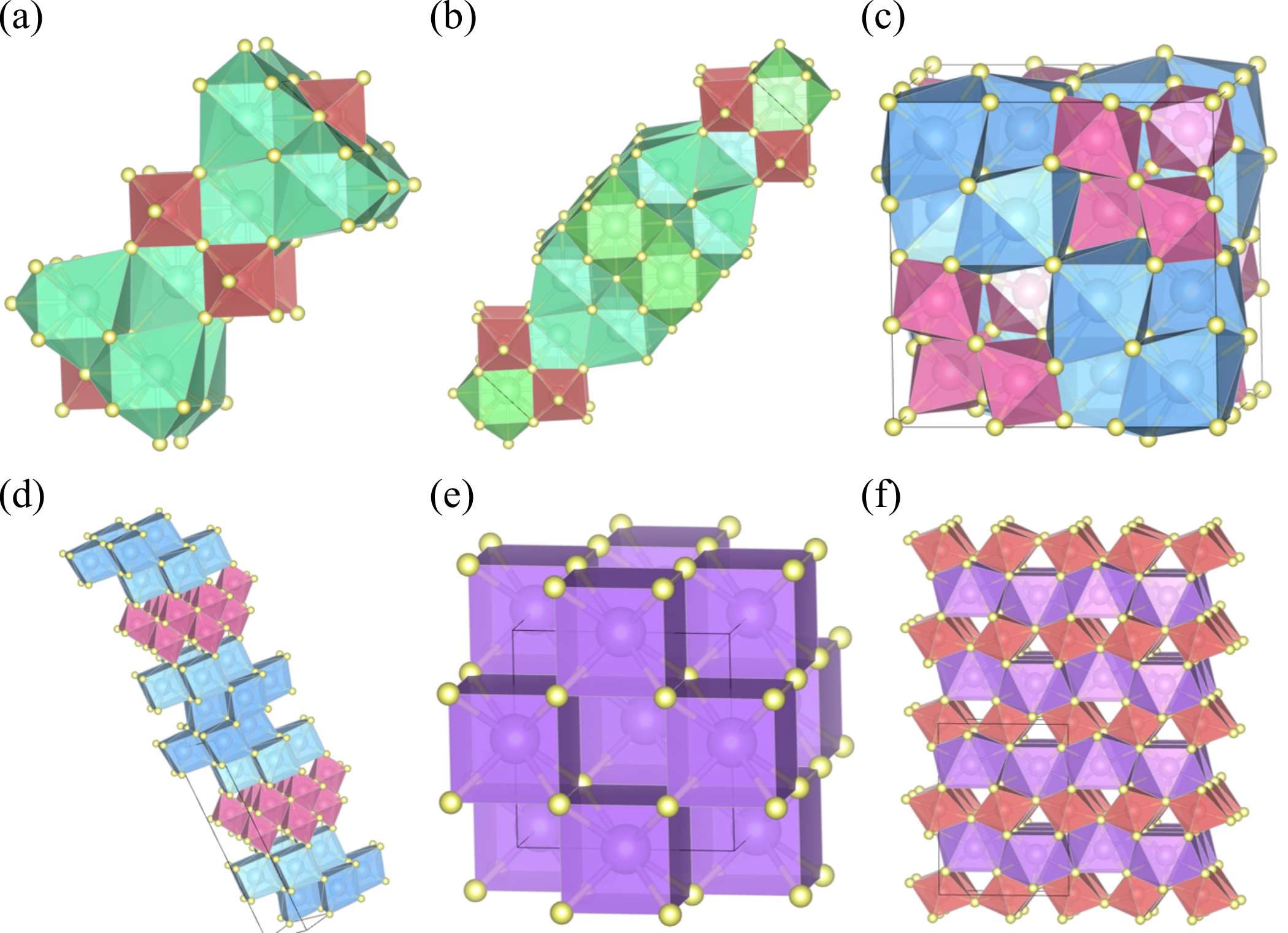}
\caption{
(a) $\mathrm{De}_8 \mathrm{Ht}_4 \mathrm{O}_{16}$ corresponding to the crystal of $\mathrm{La}_{2-x} \mathrm{Sr}_x \mathrm{CuO}_4$ (4 LGCs).
(b) $\mathrm{D}_2 \mathrm{De}_4 \mathrm{Ht}_4 \mathrm{O}_{14}$ corresponding to the crystal of $\mathrm{La}_{2-x} \mathrm{Sr}_{1+x} \mathrm{Cu}_2 \mathrm{O}_7$ (6 LGCs).
(c) $\mathrm{Uo}_4 \mathrm{Hd}_4 \mathrm{O}_{14}$ corresponding to the pyrochlore-$\alpha$ structure (4 LGCs).
(d) $\mathrm{Uo}_4 \mathrm{Hd}_4 \mathrm{O}_{14}$ with the $P1$ symmetry (6 LGCs).
(e) $\mathrm{Eo}_6 \mathrm{O}_{12}$ corresponding to the zirconia structure (2 LGCs).
(f) $\mathrm{Eo}_4 \mathrm{Ht}_4 \mathrm{O}_{12}$ with the $Cmcm$ symmetry (4 LGCs).
}
\label{fig:8-and-9_coordination}
\end{figure}

Our results indicate that $\mathrm{De}_4 \mathrm{Ht}_2 \mathrm{O}_{8}$ and $\mathrm{De}_8 \mathrm{Ht}_4 \mathrm{O}_{16}$ have only one optimal solution corresponding to the crystal structure of $\mathrm{La}_{2-x} \mathrm{Sr}_x \mathrm{CuO}_4$ as shown in Fig.~\ref{fig:8-and-9_coordination} (a). The cation $\mathrm{De}$ makes coordination polyhedron of capped square antiprism. Besides, our calculations indicate that $\mathrm{D}_1 \mathrm{De}_2 \mathrm{Ht}_2 \mathrm{O}_{7}$ and $\mathrm{D}_2 \mathrm{De}_4 \mathrm{Ht}_4 \mathrm{O}_{14}$ also have a unique optimal solution corresponding to the crystal structure of $\mathrm{La}_{2-x} \mathrm{Sr}_{1+x} \mathrm{Cu}_2 \mathrm{O}_7$ which consists of three kinds of coordination polyhedra as shown in Fig.~\ref{fig:8-and-9_coordination}(b).

In the pyrochlore-$\alpha$ structure found as one of the optimal solutions for $\mathrm{Do}_4 \mathrm{Hd}_4 \mathrm{O}_{14}$ and $\mathrm{Uo}_4 \mathrm{Hd}_4 \mathrm{O}_{14}$, one oxygen ion is placed in the center of the tetrahedron composed of large cations, and the distance of ionic bond must be smaller than those of the other ionic bonds, and thereby a large cation makes coordination polyhedron of distorted cubic as shown in Fig.~\ref{fig:8-and-9_coordination}(c). In fact, our calculations indicate that the pyrochlore-$\alpha$ structure is difficult to obtain without soft sphere, while both the $\mathrm{Do}_4 \mathrm{Hd}_4 \mathrm{O}_{14}$ and $\mathrm{Uo}_4 \mathrm{Hd}_4 \mathrm{O}_{14}$ have the optimal solution corresponding to the pyrochlore-$\alpha$ structure, which has the lowest number of LGCs with highest symmetry. Note that the interatomic distances can be as small as possible without attractive forces since the volume of the unit cell is minimized. The optimal solution consisting of the second lowest number of LGCs for the $\mathrm{Uo}_4 \mathrm{Hd}_4 \mathrm{O}_{14}$ is the layer-by-layer structure of cubics and octahedra with the $P1$ symmetry as shown in Fig.~\ref{fig:8-and-9_coordination}(d).

The cubic coordination can also be realized by the cation $\mathrm{Eo}$. $\mathrm{Eo}_6 \mathrm{O}_{12}$ has the optimal solution corresponding to the zirconia structure shown in Fig.~\ref{fig:8-and-9_coordination}(e). The cation $\mathrm{Eo}$ can also make the coordination polyhedra of square antiprism as shown in Fig.~\ref{fig:8-and-9_coordination}(f).

\begin{figure}
\centering
\includegraphics[width=1.0\columnwidth]{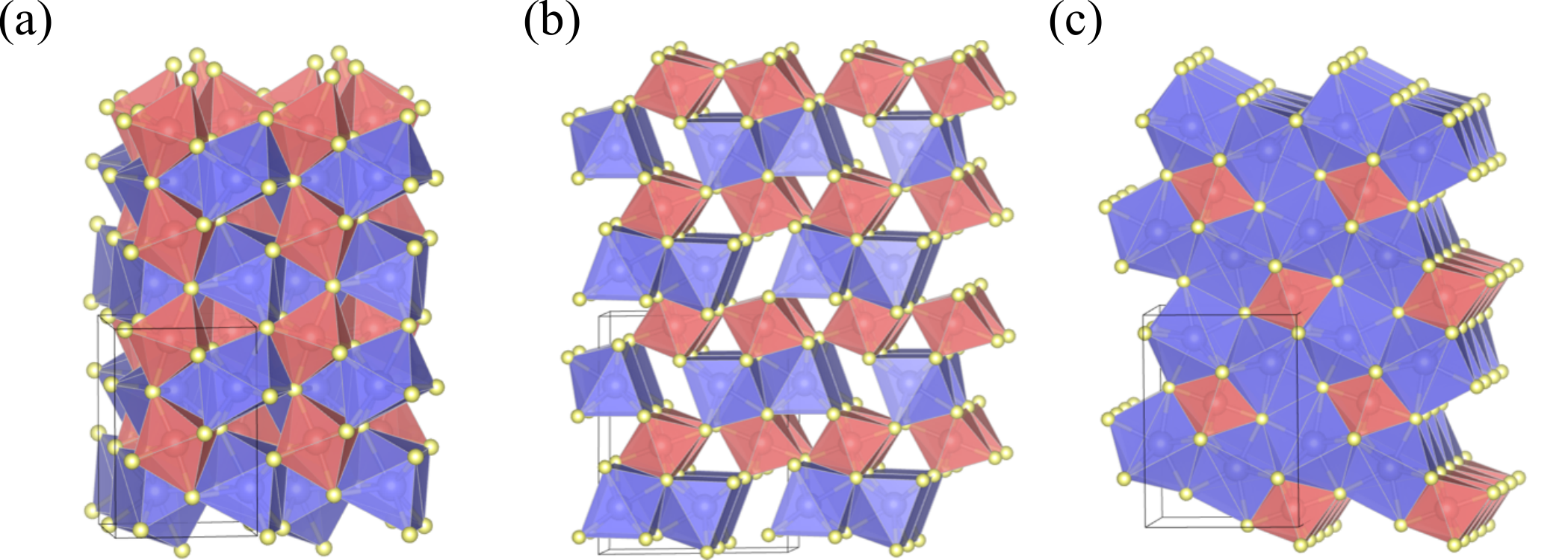}
\caption{
(a) $\mathrm{Op}_4 \mathrm{Ht}_4 \mathrm{O}_{12}$ corresponding to the crystal of $\mathrm{YFeO}_3$ (4 LGCs).
(b) $\mathrm{Op}_4 \mathrm{Ht}_4 \mathrm{O}_{12}$ with the $Pnma$ symmetry (5 LGCs).
(c) $\mathrm{Op}_4 \mathrm{Hd}_8 \mathrm{O}_{16}$ with the $Pbam$ symmetry (4 LGCs).
}
\label{fig:7_coordination}
\end{figure}

$\mathrm{Op}_4 \mathrm{Ht}_4 \mathrm{O}_{12}$ has the optimal solution corresponding to the crystal structure of $\mathrm{YFeO}_3$ which is composed of coordination polyhedra of capped trigonal prisms as shown in Fig.~\ref{fig:7_coordination}(a). $\mathrm{YFeO}_3$ and the optimal solution shown in Fig.~\ref{fig:7_coordination}(b) have the same symmetry, but the latter structure is composed of 5 LGCs, and capped trigonal prisms in the structure share the faces. The manner of atomic distribution in $\mathrm{YFeO}_3$ is the same as $\mathrm{LaMg}_{x} \mathrm{Ta}_{1-x} \mathrm{O}_{1+3x} N_{2-3x}$ which can be employed for overall water splitting at wavelengths of up to 600 nm~\cite{https://doi.org/10.1002/anie.201410961}. The correspondence implies the applicability of our method to mixed anion compounds. On the other hand, we could not find the optimal solution corresponding to the crystal structure of $\mathrm{Sr}_2 \mathrm{PdO}_4$ in the optimal solutions of $\mathrm{Op}_4 \mathrm{Hd}_8 \mathrm{O}_{16}$. However, the optimal solution, which has 4 LGCs and the $Pbam$ symmetry shown in Fig.~\ref{fig:7_coordination}(c), is similar to the crystal structure of $\mathrm{Sr}_2 \mathrm{PdO}_4$; they have the same symmetry, and the difference only comes from the kind of linking between octahedra and capped trigonal prisms. In the optimal solution, octahedra and capped trigonal prisms share the faces, while in the crystal structure of $\mathrm{Sr}_2 \mathrm{PdO}_4$, they share edges. In general, our algorithm tends to generate polyhedra sharing as many common bridging atoms as possible.

Finally, $\mathrm{Pe}_2 \mathrm{Sh}_1 \mathrm{O}_{4}$ has the optimal solution corresponding to the crystal structure of $\mathrm{InGaZnO}_{4}$. However, it may be difficult to form trigonal bipyramidal coordination polyhedra using the cation $\mathrm{Pe}$, which possesses a hard cationic radius of $C_5$. Note that in the hexagonal closest packing of oxygen ions, the center of trigonal bipyramid is identical with the interstice between three atoms in the hexagonal layer, and accordingly, the axial atoms of the bipyramid are 41\% more distant than the equatorial atoms from the central atoms. Also, note that $C_5$ is same as $C_6$.

In summary, we propose a novel method to enumerate crystal structure prototypes using the paired mathematical programming, subject to a set of constraints on atomic distances, and demonstrate its applicability to find a broad range of crystal structures of oxides. The method consists of two optimization problems. The first is the minimization problem of the volume of the unit cell under the geometrical constraints that are the minimum and maximum distances between every pair of atoms, while the second is the maximization problem of the number of chemical bonds. The constraints of the two problems make every optimal solution satisfy a series of empirical rules systematized in inorganic structural chemistry~\cite{InorganicStructuralChemistry, StructuralInorganicChemistry}. The two optimization problems are solved alternately to find a proper structure where every cation satisfies its bonding needs by transpositions of geometrical constraints as \textit{sillium} approach for amorphous silicons~\cite{WOS:000188444600058, WOS:A1985AED8200015, WOS:A1987M258200001, WOS:000089003200034}. We find that the linear relaxations of inequality constraints are effective, and accordingly, the small computational cost enables the exhaustive search for the optimal solutions of large-scale systems. We apply the mathematical programming to cases with 18 compositions, find the optimal solutions for each case successfully, and identify the corresponding real crystal structures. Our result strongly implies that the number of optimal solutions in the mathematical programming seems to be much smaller than the number of local minima in the total energy by \textit{ab-initio} simulations. A small number of optimal solutions, identified through the mathematical programming, can be easily validated using \textit{ab-initio} simulations to assess their stability. We anticipate that these optimal solutions may lead to the discovery of novel materials. The successful reproduction of a broad range of oxide crystal structures may suggest the emergence of a universal rule, potentially resulting from the precise refinement of the Pauling rules as demonstrated in this study. Given the proven success with oxide cases, it is anticipated that this method could be extended to other cases such as chalcogenides, mixed anionic compounds, intermetallic compounds, and borides, provided that effective principles are established.

R. K. is financially supported by the Grant-in-Aid for JSPS Research Fellow.
This research was conducted using FUJITSU Server PRIMERGY GX2570 (Wisteria/BDEC-01) at the Information Technology Center, The University of Tokyo, and the facilities (supercomputer Ohtaka) of the Supercomputer Center, the Institute for Solid State Physics, the University of Tokyo.
The authors would like to thank Dr.~Kengo Nishio introducing continuous random networks.
The authors also would like to thank Profs.~Yoshihiko Okamoto, Jun-ichi Yamaura, and Zenji Hiroi for kindly discussing on inorganic structural chemistry.

\newpage

\setcounter{equation}{0}
\setcounter{table}{0}
\setcounter{figure}{0}

\renewcommand{\theequation}{S\arabic{equation}}
\renewcommand{\thetable}{S-\Roman{table}}
\renewcommand{\thefigure}{S\arabic{figure}}

\onecolumngrid

\begin{center}

\Large \textbf{Supplementary Information for Mathematical Crystal Chemistry}

\vspace{0.05\columnwidth}

\end{center}

\twocolumngrid

\section{Methods}

Our mathematical programming formulation is based on inorganic structural chemistry that describes crystal sructures. As references for our basic idea, see the textbooks [U. M\"{u}ller, \textit{Inorganic Structural Chemistry, Second Edition} (John Wiley \& Sons, Ltd, 2007)] and/or [A. F. Wells, \textit{Structural Inorganic Chemistry, Fifth Edition} (Oxford University Press, 1984)].

\subsection{Geometrical constraints}

\begin{figure*}
\centering
\includegraphics[width=2.0\columnwidth]{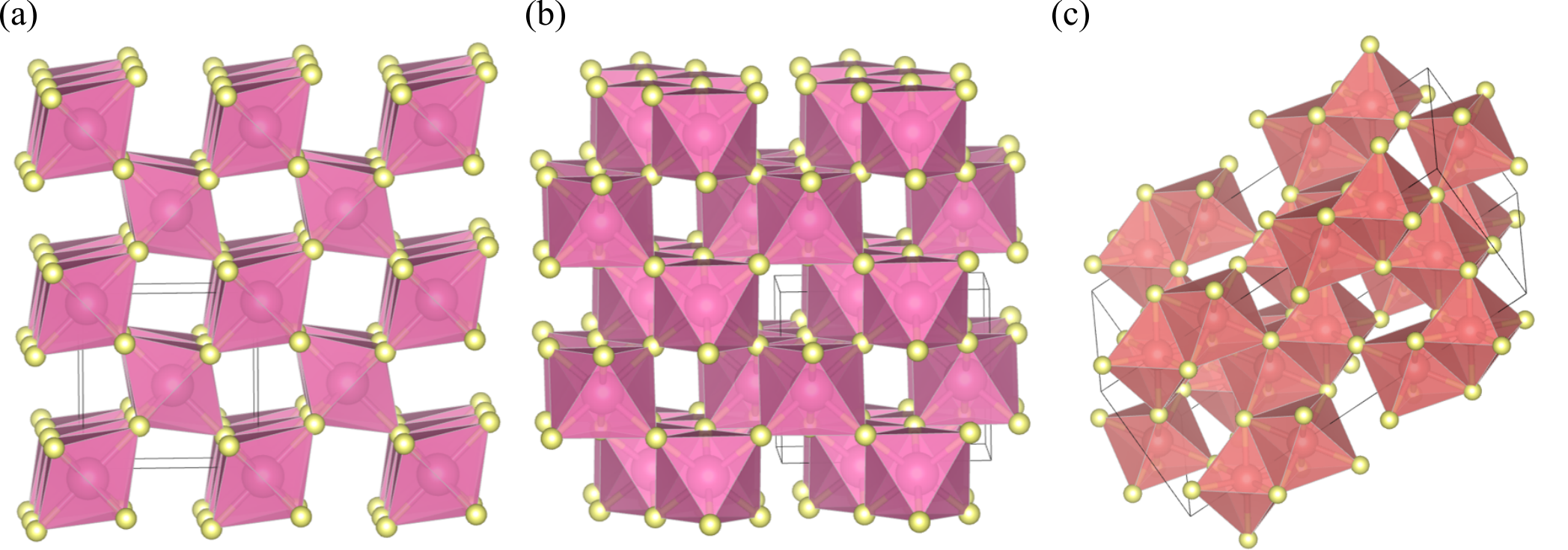}
\caption{
(a) The optimal solution of $\mathrm{Hd}_4 \mathrm{O}_8$ corresponding to the rutile structure. The space group is $P4_2/mnm$.
(b) The optimal solution of $\mathrm{Hd}_4 \mathrm{O}_8$ corresponding to the $\alpha$-$\mathrm{PbO}_2$ structure. The space group is $Pbcn$.
(c) The optimal solution of $\mathrm{Ht}_4 \mathrm{O}_6$ corresponding to the corundum structure. The space group is $R \bar{3} c$.
}
\label{fig:Titanium_Oxides}
\end{figure*}

Figure \ref{fig:Titanium_Oxides}(a) shows the optimal solution of $\mathrm{Hd}_4 \mathrm{O}_8$ correponding to the rutile structure. Every oxygen makes three chemical bonds with $\mathrm{Hd}$, while every $\mathrm{Hd}$ makes six chemical bonds with oxygens. Parallel linear strands of edge-sharing octahedra are joined by common octahedron vertices. Oxygen ions constitute the hexagonal closest-packings, and the structure is reproduced by the anionic constraints between every pair of oxygen ions. Figure \ref{fig:Titanium_Oxides}(b) shows the optimal solution of $\mathrm{Hd}_4 \mathrm{O}_8$ correponding to the $\alpha$-$\mathrm{PbO}_2$ structure. The zigzag chains of edge-sharing octahedra are also joined by common vertices. If a pair of oxygen and $\mathrm{Hd}$ is not connected by a chemical bond, a non-bonding constraint connects them. Every $\mathrm{Hd}$ makes ten polyhedral constraints, and two of them correspond to edge-sharing. These two optimal solutions, which correspond to the rutile structure and the $\alpha$-$\mathrm{PbO}_2$ structure, respectively, have the same structural fingerprint, because the two structures have the same number of LGCs of the same type, where a fingerprint of a LGC enumerates the element symbols of the adjacent atoms linked by chemical bonds, polyhedral constraints, and anionic constraints in dictionary order, respectively. Figures \ref{fig:Titanium_Oxides}(c) shows the optimal solution $\mathrm{Ht}_4 \mathrm{O}_6$ correponding to the corundum structure. There are pairs of face-sharing octahedra, and every octahedron shares three edges within a layer and three vertices with octahedra from the adjacent layer to which it has no face-sharing connection.

To share edges or faces of coordination polyhedra, the mimimum distance between two cations is shortened compared to that of the cationic constraints. Suppose $\sigma_i$ is the minimum bonding distance given by
\begin{equation}
\sigma_i = r_i ^{\left(\mathrm{I} \right)} + C_{O},
\end{equation}
and then the minimum distance $D ^{\left(\mathrm{shared} \right)}$ from the cation $i$ to the shared common edge or face is given by
\begin{equation}
D_i ^{\left(\mathrm{shared} \right)} =
\begin{cases}
\sqrt{\sigma_i ^2 - C_{O}^2} & \text{$N_{ij} ^{\left( \mathrm{CBA} \right)} = 2$} \\
\displaystyle \sqrt{\sigma_i ^2 - \frac{4}{3} C_{O}^2} & \text{$N_{ij} ^{\left( \mathrm{CBA} \right)} = 3$} \\
\sqrt{\sigma_i ^2 - 2 C_{O}^2} & \text{$N_{ij} ^{\left( \mathrm{CBA} \right)} = 4$}
\end{cases}.
\end{equation}

\subsection{Initial structures}

Initial structures are generated as follows: Suppose $\bm{a}_1$, $\bm{a}_2$ and $\bm{a}_3$ are the primitive lattice vectors. First, they are given by
\begin{align}
\left(\bm{a}_1 , \bm{a}_2 , \bm{a}_3 \right) = \begin{pmatrix}
l_1 && l_2 \cos \theta_2 && l_3 \sin \varphi_3 \cos \theta_3 \\
0 && l_2 \sin \theta_2 && l_3 \sin \varphi_3 \sin \theta_3 \\
0 && 0 && l_3 \cos \varphi_3
 \end{pmatrix},
\end{align}
where $1 \le l_i \le 2$, $\displaystyle \frac{\pi}{3} \le \theta_i \le \frac{2 \pi}{3}$, and $\displaystyle - \frac{\pi}{6} \le \varphi_i \le \frac{\pi}{6}$ are randomly set. Second, the lattice is expanded until the sum of the volume of the atomic spheres becomes 70\% of the volume of the unit cell, where the anionic and cationic radii are set to be $0.6$ and $1.4$, respectively. The cartesian coordinate of an atom is set to be
\begin{equation}
\bm{x} = q_1 \bm{a}_1 + q_2 \bm{a}_2 + q_3 \bm{a}_3
\end{equation}
where $0 \le q_i \le 1$ are random values.

\subsection{Cationic Radii}

\begin{table}
\caption{List of cationic radii. $C_n$ ($n \in \mathbb{N}$) corresponds to the minimum radius so that the cation can connect $n$ oxygen ions, and $C_{O}$ correponds to the ionic radius of oxygen ion.}
\label{table:cationicRadii}
\begin{ruledtabular}
\begin{tabular}{cc}
Symbol & Cationic radius \\ \hline
$C_4$ & $0.314643$ \\
$C_5$ & $0.579899$ \\
$C_6$ & $0.579899$ \\
$C_7$ & $0.827755$ \\
$C_8$ & $0.903460$ \\
$C_9$ & $1.02487$ \\
$C_{10}$ & $1.165450$ \\
$C_O$ & $1.4$ \\
\end{tabular}
\end{ruledtabular}
\end{table}

The $11$ cations, which are listed in Table II of the main article, are selected by considering closed packed polyhedra consisting of the centered cation and surrounding oxygen ions. Cationic radius $C_n$ ($n \in \mathbb{N}$) corresponds to the minimum radius so that the cation can connect $n$ oxygen ions, while $C_{O}$ correponds to the ionic radius of oxygen ion. Table \ref{table:cationicRadii} lists the value of $C_n$ and $C_{O}$.

All cations have the same cationic repulsion radius $r^{\left( \mathrm{C} \right)} = 1.4$ which is the same as the ionic radius of oxygen ion to avoid the condensation of cations.

\section{Numerical aspects}

\begin{table}
\caption{Common parameters for global and local geometrical optimizations.}
\label{table:common_parameters_for_geometrical_optimization}
\begin{ruledtabular}
\begin{tabular}{cc}
Parameter & Value \\ \hline
Pressure & $1.0$ \\
Attractive force constant $k_{\uparrow}$ & $30.0$ \\
Repulsive force constant $k_{\downarrow}$ & $-100.0$ \\
Maximum error of geometrical constraints & $5 \%$ \\
Minimum packing fraction & $60 \%$ \\
Maximum number of total optimization steps & 40000
\end{tabular}
\end{ruledtabular}
\end{table}
\begin{table}
\caption{Specific parameters for global geometrical optimization.}
\label{table:specific_parameters_for_global_geometrical_optimization}
\begin{ruledtabular}
\begin{tabular}{cc}
Parameter & Value \\ \hline
Initial $\Delta x_{\mathrm{max}}$ & $6.0$ \\
Final $\Delta x_{\mathrm{max}}$ & $0.55$ \\
Maximum number of optimization steps & $40000$ \\
Update frequency of geometrical constraints & $20$
\end{tabular}
\end{ruledtabular}
\end{table}
\begin{table}
\caption{Specific parameters for local geometrical optimization. The maximum displacement is gradually decreased to find optimal solutions.}
\label{table:specific_parameters_for_local_geometrical_optimization}
\begin{ruledtabular}
\begin{tabular}{cc}
Parameter & Value \\ \hline
Initial $\Delta x_{\mathrm{max}}$ & $0.1$ \\
Final $\Delta x_{\mathrm{max}}$ & $0.02$ \\
Maximum number of optimization steps & $1000$
\end{tabular}
\end{ruledtabular}
\end{table}

Table \ref{table:common_parameters_for_geometrical_optimization} lists the common parameters for the global and local geometrical optimizations. The geometrical constraints are given by the minimum and maximum distances between atoms as
\begin{align}
x_{ij \bm{T}} \ge d_{ij \bm{T}} ^{\left( \mathrm{min} \right)}, && x_{ij \bm{T}} \le d_{ij \bm{T}} ^{\left( \mathrm{max} \right)},
\end{align}
but when we confirm the feasibility of the structure after local geometrical optimization, we permit $5\%$ error as
\begin{align}
x_{ij \bm{T}} \ge 0.95 \, d_{ij \bm{T}} ^{\left( \mathrm{min} \right)}, && x_{ij \bm{T}} \le 1.05 \, d_{ij \bm{T}} ^{\left( \mathrm{max} \right)}.
\label{eq:geometrical_constraints_with_error}
\end{align}
Besides, in this study, if the packing fraction is less than the minimum, we reject the sparse structure from the optimum solutions. $\alpha$ is calculated as
\begin{equation}
\alpha = \left( \frac{\Delta x_{\mathrm{max, f}}}{\Delta x_{\mathrm{max, i}}} \right) ^{\frac{1}{N_{\mathrm{opt}}}},
\end{equation}
where $\Delta x_{\mathrm{max, i}}$ and $\Delta x_{\mathrm{max, f}}$ are the initial and final $\Delta x_{\mathrm{max}}$, respectively, and $N_{\mathrm{opt}}$ is the maximum number of optimization steps.

Table \ref{table:specific_parameters_for_global_geometrical_optimization} lists the specific parameters for the global geometrical optimizations (GGO). The network of the chemical bonds is updated per 20 steps of GGO. We monitor the crystal lattice per 1000 steps of GGO, and we reject the structure if the lattice is too distorted or too sparse. To reduce the computational costs, the neighboring list of pairs of atoms is updated per 60 steps of GGO.

Table \ref{table:specific_parameters_for_local_geometrical_optimization} lists the specific parameters for the local geometrical optimizations (LGO). The geometrical constraints are fixed through LGO. To relax the structure, the maximum displacements of atoms and crystal lattice are gradually decreased.

\begin{table}
\caption{The number of samples reaching optimal solutions.
We generate $1000000$ samples for $\mathrm{Ht}_4 \mathrm{T}_2 \mathrm{O}_8$ and $\mathrm{Ht}_8 \mathrm{T}_4 \mathrm{O}_{16}$.
Note that $746250$ samples of $\mathrm{Ht}_4 \mathrm{T}_2 \mathrm{O}_{8}$ and $938542$ samples of $\mathrm{Ht}_8 \mathrm{T}_4 \mathrm{O}_{16}$ cannot reach optimal solutions, respectively.
}
\label{table:number of samples}
\begin{ruledtabular}
\begin{tabular}{cc}
Optimal Solution & Number of samples \\ \hline
$\mathrm{Ht}_4 \mathrm{T}_2 \mathrm{O}_8$ (spinel structure) & $126048$ \\
$\mathrm{Ht}_4 \mathrm{T}_2 \mathrm{O}_8$ ($Cmmm$ symmetry) & $6061$ \\
$\mathrm{Ht}_8 \mathrm{T}_4 \mathrm{O}_{16}$ (spinel structure) & $906$ \\
$\mathrm{Ht}_8 \mathrm{T}_4 \mathrm{O}_{16}$ ($Imma$ symmetry) & $800$
\end{tabular}
\end{ruledtabular}
\end{table}

\begin{figure*}
\centering
\includegraphics[width=2.0\columnwidth]{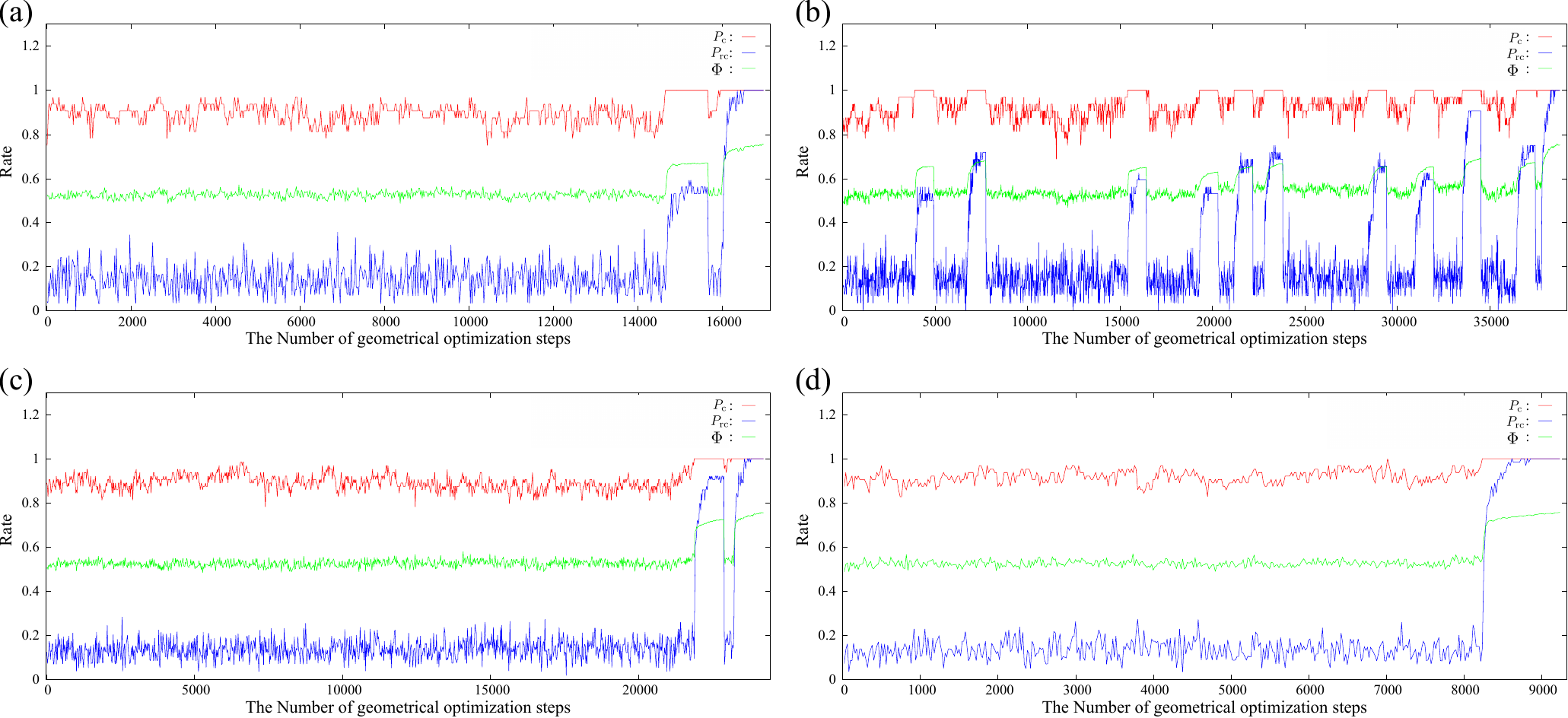}
\caption{
The examples of the changes of the rate of chemical bonds and relaxed chemical bonds, and the packing fraction through the optimization processes.
(a) The exaples of $\mathrm{Ht}_4 \mathrm{T}_2 \mathrm{O}_8$ that is optimized to the spinel structure. In the first local optimization process, $P_{\mathrm{rc}}$ cannot reach $1.0$.
(b) The exaples of $\mathrm{Ht}_4 \mathrm{T}_2 \mathrm{O}_8$ that is optimized to the structure with the $Cmmm$ symmetry. This structure tends to require many local optimization processes.
(c) The exaples of $\mathrm{Ht}_8 \mathrm{T}_4 \mathrm{O}_{16}$ that is optimized to the spinel structure. In the first local optimization process, $P_{\mathrm{rc}}$ cannot reach $1.0$.
(d) The exaples of $\mathrm{Ht}_8 \mathrm{T}_4 \mathrm{O}_{16}$ that is optimized to the structure with the $Imma$ symmetry. This sample reaches the optimal solution in the first local optimization process.
}
\label{fig:chemicalBondingRecord}
\end{figure*}

We generate $1000000$ samples for $\mathrm{Ht}_4 \mathrm{T}_2 \mathrm{O}_8$ and $\mathrm{Ht}_8 \mathrm{T}_4 \mathrm{O}_{16}$. Table \ref{table:number of samples} shows the number of $\mathrm{Ht}_4 \mathrm{T}_2 \mathrm{O}_8$ and $\mathrm{Ht}_8 \mathrm{T}_4 \mathrm{O}_{16}$ samples optimized to the spinel structure; about $10 \%$ of $\mathrm{Ht}_4 \mathrm{T}_2 \mathrm{O}_8$ are optimized to spinel structure, while $0.01 \%$ of $\mathrm{Ht}_8 \mathrm{T}_4 \mathrm{O}_{16}$ are optimized to spinel structure. $938542$ samples of $\mathrm{Ht}_8 \mathrm{T}_4 \mathrm{O}_{16}$ cannot reach optimal solutions if we use the optimization parameters shown in Tables \ref{table:common_parameters_for_geometrical_optimization}, \ref{table:specific_parameters_for_global_geometrical_optimization}, and \ref{table:specific_parameters_for_local_geometrical_optimization}.

Generally, not all the cations have the maximum number of chemical bonds. The global geometrical optimization is aimed at transforming the structure largely enough to create a different network of chemical bonds so that every cation satisfies its bonding needs. To analyze the optimization history, we monitor the rate of chemical bonds $P_{\mathrm{c}}$ given by
\begin{equation}
P_{\mathrm{c}} \equiv \frac{\sum_i s_{i} ^{\left(+ \right)} n_{i} ^{\left(\mathrm{CB} \right)}}{\sum_i s_{i} ^{\left(+ \right)} N_{i} ^{\left(\mathrm{CB} \right)}},
\end{equation}
where the numerator is the number of chemical bonds, and the denominator is the total number of bonding needs. If every cation satisfies its bonding needs, the structure is locally optimized to identify whether the structure can be the optimal solusion of the geometrical optimization problem. If so, all the interatomic distances satisfy the condition of Eq.~\eqref{eq:geometrical_constraints_with_error}, and we regard the chemical bonds satisfying Eq.~\eqref{eq:geometrical_constraints_with_error} as the relaxed chemical bonds. We monitor the rate of relaxed chemical bonds $P_{\mathrm{rc}}$, where the numerator and the denominator are the total number of the relaxed chemical bonds and the total number of the chemical bonds of the constraint, respectively. We also monitor the packing fraction $\Phi$ defined to be
\begin{equation}
\Phi = \frac{1}{\Omega} \sum_i \frac{4 \pi}{3} \left(r_i ^{\left(\mathrm{I} \right)} \right)^3.
\end{equation}
Figure \ref{fig:chemicalBondingRecord} shows the changes of $P_{\mathrm{c}}$, $P_{\mathrm{rc}}$ and $\Phi$ during the optimization process. The fluctuation of $P_{\mathrm{c}}$ indicates that a structure is transformed largely enough to change the network of chemical bonds. The small rate of $P_{\mathrm{rc}}$ comes from the large displacement through the global optimization. If $P_{\mathrm{c}}$ reaches $1.0$, the structure is locally optimized to identify whether the structure can be the optimal solution of the geometrical optimization problem within the $5 \%$ error. In local optimization, $P_{\mathrm{rc}}$ shows larger value due to the small displacement, however, in some cases, $P_{\mathrm{rc}}$ cannot reach $1.0$ due to the contradictions in the geometrical constraints. $\Phi$ has the value around $0.5$ through the global optimization process.

\end{document}